\documentclass[pdflatex,sn-mathphys]{sn-jnl}% Math and Physical Sciences

\theoremstyle{thmstyleone}%
\theoremstyle{thmstyletwo}%

\theoremstyle{thmstylethree}%
\jyear{2023}%

\begin{document}

%\title{Learning the Front Dynamics of Phase Field Systems:\\
%Black and Gray Box Surrogates}
\title[ ]{Physics-agnostic and Physics-infused machine learning for thin films flows:
modeling, and predictions from small data}

% \title{Physics-agnostic and Physics-infused machine learning for thin films flows: \\
% modeling, and predictions from small  data}
%From Navier-Stokes data to thin film amplitude equations: data-driven strategies for learning interface equations and models for out-of-sample predictions from gappy data}

\author[1]{\fnm{Cristina P.} \sur{Martin-Linares}}\email{cristina.martinlinares@jhu.edu}

\author[2]{\fnm{Yorgos M.} \sur{Psarellis}}\email{gpsarel1@jhu.edu}

\author[3]{\fnm{Georgios} \sur{Karapetsas}}\email{gkarapetsas@auth.gr}
\author*[4]{\fnm{Eleni D.} \sur{Koronaki}}\email{eleni.koronaki@uni.lu}

\author*[2]{\fnm{Ioannis G.} \sur{Kevrekidis}}\email{yannisk@jhu.edu}

\affil[1]{\orgdiv{Department of Mechanical Engineering}, \orgname{Whiting School of Engineering, Johns Hopkins University}, \orgaddress{\street{3400 North Charles Street}, \city{Baltimore}, \postcode{21218}, \state{MD}, \country{USA}}}

\affil*[2]{\orgdiv{Department of Chemical and Biomolecular Engineering}, \orgname{Whiting School of Engineering, Johns Hopkins University}, \orgaddress{\street{3400 North Charles Street}, \city{Baltimore}, \postcode{3400}, \state{MD}, \country{USA}}}

\affil[3]{\orgdiv{Department of Chemical Engineering}, \orgname{Aristotle University of Thessaloniki}, \orgaddress{ \city{Thessaloniki}, \postcode{54124}, \country{Greece}}}

\affil*[4]{\orgdiv{Interdisciplinary Center for Security, Reliability and Trust}, \orgname{University of Luxembourg}, \orgaddress{\street{29 John F. Kennedy Avenue}, \city{Luxembourg}, \postcode{1855}, \country{Luxembourg}}}

% \author{Cristina P. Martin-Linares}
% %\email[]{cristina.martinlinares@jhu.edu}
% \affiliation{
%   Department of Mechanical Engineering,
%   Whiting School of Engineering, Johns Hopkins University,
%   3400 North Charles Street, Baltimore, MD 21218, USA
% }

% \author{Yorgos M. Psarellis}
% \affiliation{
%   Department of Chemical and Biomolecular Engineering,
%   Whiting School of Engineering, Johns Hopkins University,
%   3400 North Charles Street, Baltimore, MD 21218, USA
% }

% \author{Georgios Karapetsas}
% \affiliation{
%   Department of Chemical Engineering,
%   Aristotle University of Thessaloniki,
%   54124 Thessaloniki, GREECE
% }

% \author{Eleni D. Koronaki}
% \email[]{eleni.koronaki@uni.lu}
% %\thanks{authors contributed equally}
% \affiliation{
%   Interdisciplinary Center for Security, Reliability and Trust,
%   University of Luxembourg,
%   29 John F. Kennedy Avenue 1855 Luxembourg, LUXEMBOURG
% }

% \author{Ioannis G. Kevrekidis}
% \email[]{yannisk@jhu.edu}
% \affiliation{
%   Department of Chemical and Biomolecular Engineering,
%   Whiting School of Engineering, Johns Hopkins University,
%   3400 North Charles Street, Baltimore, MD 21218, USA
% }

\date{\today}

%%%YGK   the unstable branch in the bif diagram
%%%YGK    the "right references" to deriving the KS  and the "higher order equations" like Benney

\abstract{Numerical simulations of multiphase flows are crucial in numerous engineering applications, but are often limited by the computationally demanding solution of the Navier-Stokes (NS) equations. Here, we present a data-driven workflow where a handful of detailed NS simulation data are leveraged into a reduced-order model for a prototypical vertically falling liquid film. We develop a physics-agnostic model for the film thickness, achieving a far better agreement with the NS solutions than the asymptotic Kuramoto-Sivashinsky (KS) equation. We also develop two variants of physics-infused models providing a form of calibration of a low-fidelity model (i.e. the KS) against a few high-fidelity NS data. Finally, predictive models for missing data are developed, for either the amplitude, or the full-field velocity and even the flow parameter from partial information. This is achieved with the so-called "Gappy Diffusion Maps", which we compare favorably to its linear counterpart, Gappy POD.}

\maketitle

\section{Introduction}\label{sec1}
% \section{Introduction}
The study of multiphase flows is often limited by the computational effort involved in solving the Navier-Stokes equations \cite{glasser1997fully}. One such example, the flow of thin films of liquid on inclined planes, has fascinated researchers not only because of the wide range of industrial applications but also because of the interesting dynamics of the liquid-air interface \cite{Kalliadasis2012}. The Navier-Stokes (NS) equations accurately describe the fluid motion and also the evolution of the surface but suffer from high computational cost \cite{Pettas2019a,Pettas2019b}.
To this end, significant effort has led to several approximate interface evolution equations that are much simpler to solve but are nevertheless valid under specific assumptions and limitations. Beyond their limits of validity, it is often found that they yield nonphysical solutions, or even blow up \cite{Kalliadasis2012}, posing significant restrictions to their applicability.

In order to drastically enable Computational Fluid Dynamics and break new barriers in flow control, uncertainty quantification and shape optimization, it is crucial to develop novel, robust and efficient data-driven/data-assisted models that combine physical and mathematical insight with machine learning strategies.
 This work presents a methodology for deriving data-driven partial differential equations (PDEs) for the film amplitude, based on a collection of NS simulation data, that are not subject to restrictions and assumptions for the flow and hence are more general. Our work falls in the category of dynamical system identification \cite{gonzalez-garcia98_ident_distr_param_system, krischer93_model_ident_spatiot_varyin_catal_react, rico-martinez92_discr_vs,kemeth22_learn_emerg_partial_differ_equat}. 
 Recently increased interest in PDE identification has led to the development of alternative algorithmic tools, such as sparse identification of nonlinear dynamical systems using dictionaries~\cite{brunton16_discov_gover_equat_from_data, rudy17_data_driven_discov_partial_differ_equat}, PDE-net~\cite{long17_pde_net}, physics-informed neural networks~\cite{raissi19_physic_infor_neural_networ}, and others \cite{Chen2018, Vlachas2018, Vlachas2022}
 Our algorithmic approach can be implemented on data from detailed PDE simulations \cite{Psarellis2022_2}, agent-based modeling \cite{Psarellis2022, Arbabi2021} or Lattice Boltzmann simulations \cite{Galaris2022, Lee2020} among others. Extensions of PDE identification including gray-box or closure identification (such as those explored in our work) have been studied in the context of various applications \cite{Duraisamy2015, Duraisamy2019, Psarellis2022, Psarellis2022_2, Lee2019, Kemeth2022, Pan2018, PARISH2016758}. In the relevant literature, the Kuramoto-Sivashinsky (KS) equation, selected in this work as a low-fidelity counterpart of the NS equations,  has served as a benchmark case study, due to its wealth of dynamic responses and highly nonlinear nature \cite{gonzalez-garcia98_ident_distr_param_system,   Raissi2017, Vlachas2018,linot2023stabilized,floryan2022data}.

%%%YGK
%%  both the additive and the functional are Gray boxes
%% one is an additive gray box
%% and the other a functional gray box
%%
%%   we need more references on closures...
%%
%%%
The results of the learned PDE are compared to the ground-truth NS results and also the results of the KS equation. It is expected that past a certain limit, the KS will perform poorly and produce nonphysical solutions. Yet it is still useful in the context of learning an accurate amplitude PDE, as it will be shown, in two different ways: In the first "Gray box" model approach, an \textit{additive correction} of the KS, can be learned from NS data as a sort of calibration of the low-fidelity model against high-fidelity data. In this context, the data-driven model provides a measure of the discrepancy between the approximate equation and the ground truth, and serves to inform as to the actual limits of applicability of the KS in terms of the flow parameter, here the Reynolds number. In the second approach, which we call the "functional correction Gray Box model", certain observations of the KS formula, such as the value of its right-hand-side, its derivatives or even values in specific nearby time-instances or nearby points in space, are used as inputs to the learned model. 

In addition, reduced representations of the NS data, including full velocity fields and fluid film height, are further exploited for out-of-sample predictions from partial data at the small data limit. Nonlinear manifold learning, specifically Diffusion Maps and linear methods, i.e. Proper Orthogonal Decomposition (POD) are initially implemented, in order to derive a low-order description of the high-dimensional data-set. It is then shown that efficient interpolation in the reduced space can help recover entire sets of data from partial information. 
Specifically, it will be demonstrated that full velocity profiles, parameter values and film height measurements can be predicted given a handful of values for the film height at specific locations. The advantages of {\em "Gappy" Diffusion Maps} over its linear counterpart, Gappy POD, are discussed in relation to the parsimony of the description of the manifold that contains the data and to the location of the known measurements.

%In the results that will be presented below, the time dependent equations are solved in an Eulerian frame. We consider the flow over a vertical plane, i.e. $\theta=90^\circ$ with a domain of dimensionless length, $l=L/H_N=95$; $L$ denotes the dimensional length of the domain. The value of the Weber number used in the simulations is W=278. We start our simulations from an initial condition corresponding to a flat film perturbed by a sinusoidal perturbation with amplitude of 3\% of the dimensionless Nusselt film height, $H=1$. The height of the film, $h(x,t)$, is collected at each time-step until a steady travelling wave is formed.

%\section{The Kuramoto-Sivashinsky equation}
%The Kuramoto-Sivashinsky equation is a single-equation surrogate of the amplitude dynamics. Assuming the flow over a vertical plane, the KS equation can be written as 

%\begin{equation}
%    \begin{split}
%     & \phi_\tau=-\alpha(\phi \phi_\xi+\phi_{\xi\xi})-4\phi_{\xi\xi\xi\xi}; \;\; \textrm{for} \;\; \xi\in[0,2\pi]\\
%     & \alpha=\frac{8}{5} \frac{R F}{\epsilon^2 W}\\
%   \end{split}
%    \label{eq:KS}
% \end{equation}

 \section{Results}
\subsection{Bifurcation diagram: NS vs KS}
\begin{figure}[ht]
\centering
\includegraphics[width=\columnwidth]{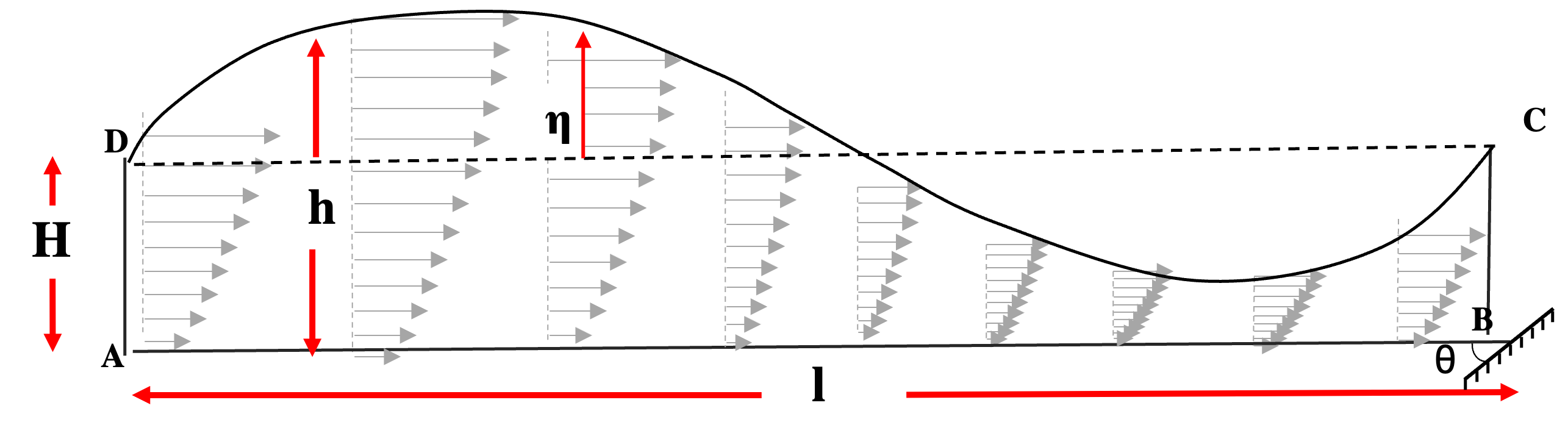}
\caption{Cross-section of a film flowing on plane, inclined with respect to the horizontal by angle $\theta$.  $H=1$ is the dimensionless Nusselt film height }
\label{fig:film}
\end{figure}

The NS data necessary for learning the amplitude equations are derived by solving the time dependent equations, as presented in Methods, in an Eulerian frame. We consider the flow over a vertical plane, schematically presented in ~\ref{fig:film}, i.e. $\theta=90^\circ$ with a domain of dimensionless length, $l=L/H_N=95$; $L$ denotes the dimensional length of the domain. The value of the Weber number used in the simulations is W=278. We start our simulations from an initial condition corresponding to a flat film perturbed by a sinusoidal perturbation with amplitude ~3\% of the dimensionless Nusselt film height, $H=1$. The height of the film, $h(x,t)$, is collected at each time-step until a steady travelling wave is formed.

 The single-equation surrogate of the amplitude, selected in this study, is the Kuramoto-Sivashinsky equation. Assuming the flow over a vertical plane, the KS equation can be written as 

\begin{equation}
   \begin{split}
    & \phi_\tau=-\alpha(\phi \phi_\xi+\phi_{\xi\xi})-4\phi_{\xi\xi\xi\xi}; \;\; \textrm{for} \;\; \xi\in[0,2\pi]\\
    & \alpha=\frac{8}{5} \frac{R F}{\epsilon^2 W}\\
  \end{split}
   \label{eq:KS}
\end{equation}

\begin{figure}[ht]
\centering
\includegraphics[width=0.5\columnwidth]{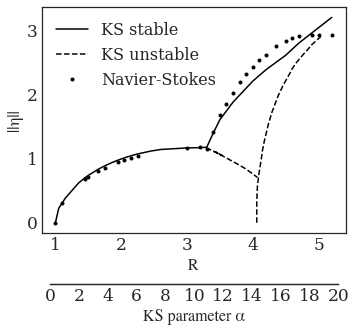}
\caption{Bifurcation diagram of the Navier-Stokes and the Kuramoto-Sivashinsky}
\label{fig:NSvsKS}
\end{figure}

The KS can be derived by the NS under certain assumptions which are summarized for completeness in Methods. It would be useful to briefly discuss the limitations of the KS equation before proceeding with the presentation of our results. 
Even though, it is well-known that the KS equation is valid for Reynolds number values of O(1), to the best of our knowledge there is no direct comparison in the literature of the KS to the NS results. To clearly present the limitations of the KS equation, here we plot in Fig. ~\ref{fig:NSvsKS} the norm
%This is shown in ~\ref{fig:NSvsKS}, where the norm 
of the amplitude distribution, $||\eta||,$ with respect to both the Reynolds number and the KS parameter $\alpha$. 

Both equations predict very similar results for $1\leq R \leq 3.4$ (or $4\leq \alpha \leq 13.43$). In this range of parameter values, the solution of the KS is a stationary wave (in a co-moving frame with speed c=-3), whereas the NS, solved in an Eulerian frame evolves into a travelling wave with a steady and unchanged shape and speed c=3.
Past that point (R=3.4 and $\alpha=13.43$), the speed of the wave becomes larger than 3 and hence the solution of the KS is also travelling (with speed c-3). Up to approximately R=4.3 ($\alpha=16.5$) the KS start to gradually deviate from the NS solution. For higher values of $\alpha$, i.e. $\alpha>16.998$, the solution of the KS is a so-called "pulsing" wave, as described in detail in ~\cite{Kevrekidis1990}. The pulsing waves oscillate between two waveforms that are $\pi$-periodic in space and are $\pi/2$ shifts of each other. Such a solution has not been reported for the NS equation.

\subsection{Black box model: Learning the PDE}
Here our goal is to use data from the NS simulations to learn a PDE of the general form:
\begin{equation}
    \frac{\partial{\eta}}{\partial t}=f\left(\eta,\eta_x,\eta_{xx},\eta_{xxx},\eta_{xxxx},R\right)
    \label{eq:black}
\end{equation}

The function $f$ is approximated by a fully connected neural network. The inputs to the neural network are the amplitude and spatial derivatives of the amplitude, as extracted from the NS simulations. Specifically, the NS model is implemented for 20 parameter values, and snapshots, i.e. time-instances of the film surface evolution are collected in equally sized time-steps ($dt=1$ is the dimensionless time unit). The spatial derivatives, up to 4th order are computed using Fourier transforms, in each point is space and time. The time derivative of the amplitude can be extracted directly from the NS code (although it can also be easily computed, e.g with finite differences). 

\begin{figure}[h!]
\centering
\includegraphics[width=\columnwidth]{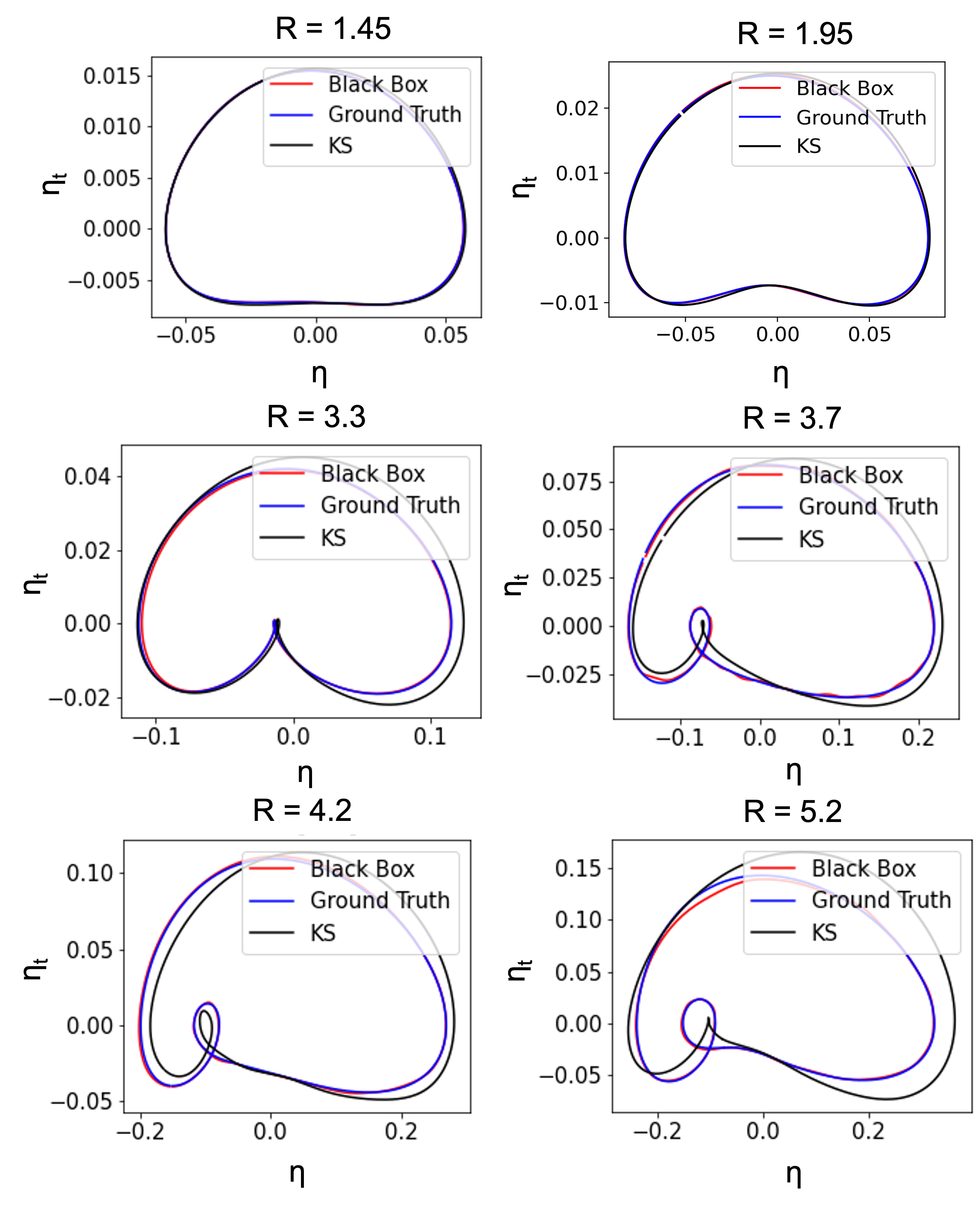}
\caption{Attractors for different R numbers derived by (i)the Navier-Stokes (shown in blue) (ii) the Kuramoto-Sivashinsky (shown in black) and (iii) the NN-derived PDE (red); all results are rescaled in the NS scaling.} 
\label{fig:Attractors}
\end{figure}

This collection of data is then used to train the neural network to predict the time derivative of the amplitude from the value of the amplitude and a few spatial derivatives. Once this is done, the right-hand-side of the PDE in Eq.~\ref{eq:black} can be used in conjunction with any method of integration in time, such as the Runge-Kutta. The attractors that resulted from the integration of the learned PDE are shown in Fig.~\ref{fig:Attractors}, for a representative selection of parameter values. The attractor of the neural network derived PDE, shown in red, is almost a perfect match with the ground-truth results of the NS (blue line). 

\begin{figure}[h!]
\centering
\includegraphics[width=\columnwidth]{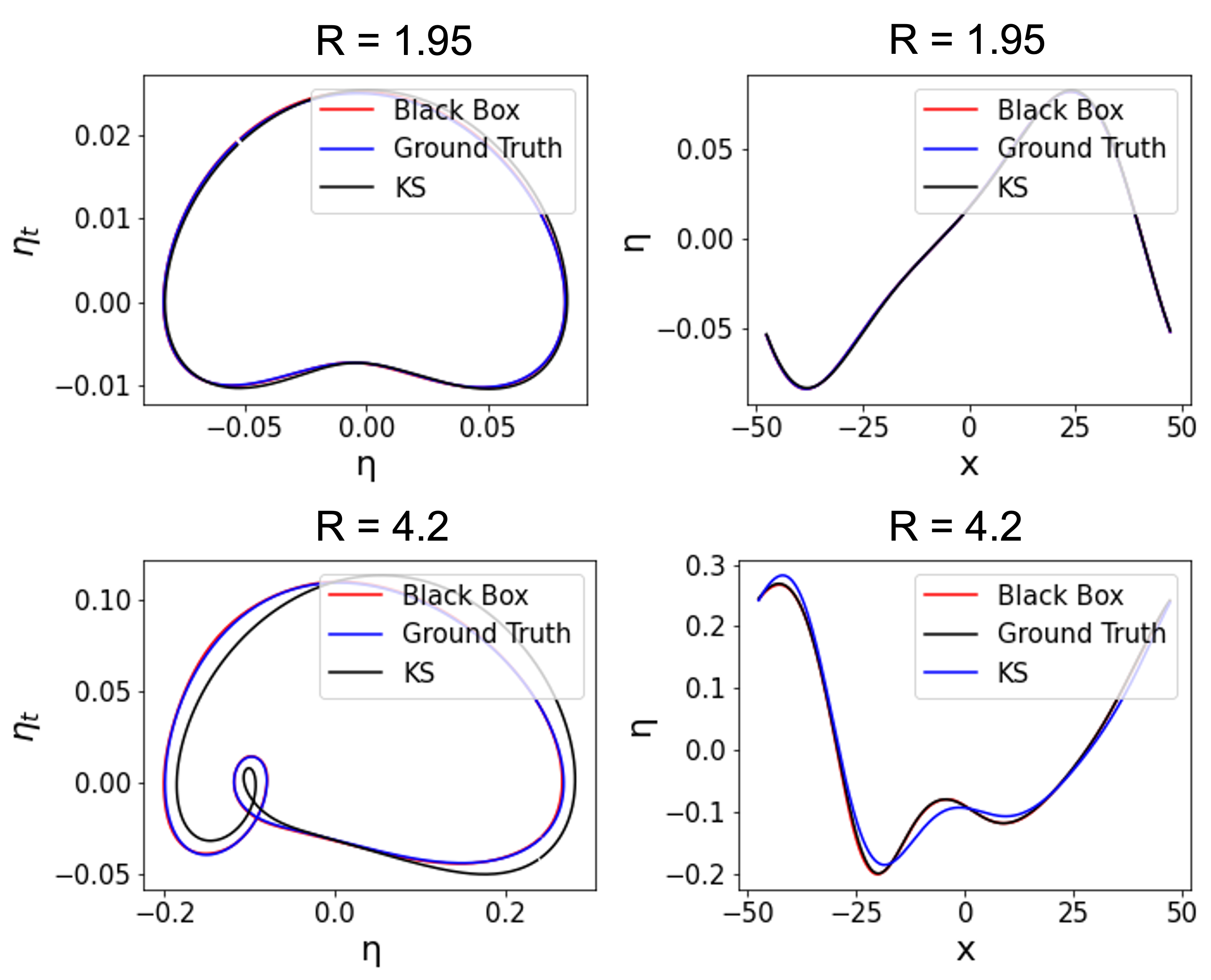}
\caption{Left:Attractors Right: time-instance of the amplitude;  derived by (i)the Navier-Stokes (shown in blue) (ii) the Kuramoto-Sivashinsky (shown in orange) and (iii) the NN-derived PDE (black);  Top row: R=1.95; Bottom row: R=4.2; all results are rescaled in the NS scaling.} 
\label{fig:Attractors_wave}
\end{figure}

For reference and comparison, the KS results (appropriately rescaled) are shown in the same plot (black line). The KS performs well for small values of the R number (R$<$3.3), but then progressively starts to deviate quantitatively for increasing values of R. This is shown in Fig.~\ref{fig:Attractors_wave}, where a snapshot of the amplitude derived by the KS, the NS and the Black Box model are shown alongside the corresponding phase portraits. Despite the apparent failure of the KS to capture the wave dynamics accurately, it still yields \textit{qualitatively} good results. We exploit this further, to infuse physical information into the data-driven amplitude equation. This is discussed in the following paragraphs.

\subsection{Gray box model I: Learning an additive correction to the KS}
Instead of training a neural network to learn the right-hand-side of a PDE as a "Black box", i.e. without any physical intuition about the function, the KS (the approximate analytical model) is used as foundation upon which a \textit{correction} is added to make it more accurate. This correction is discovered in a data-driven way, using the same data as before (described in the previous paragraph). In this case though, the output of the neural network is not the time derivative of the amplitude but rather, the \textit{difference} between the actual and the KS time-derivative:
\begin{equation}
    \left(\frac{\partial{\eta}}{\partial t}\right)^{NS}-\left(\frac{\partial{\eta}}{\partial t}\right)^{KS}=f\left(\eta,\eta_x,\eta_{xx},\eta_{xxx},\eta_{xxxx},R\right);
    \label{eq:gray}
\end{equation}
\noindent 
this can be thought of then as  a "Residual Network", a ResNet \cite{he2015deep}.

\begin{figure}[h]
\centering
\includegraphics[width=\columnwidth]{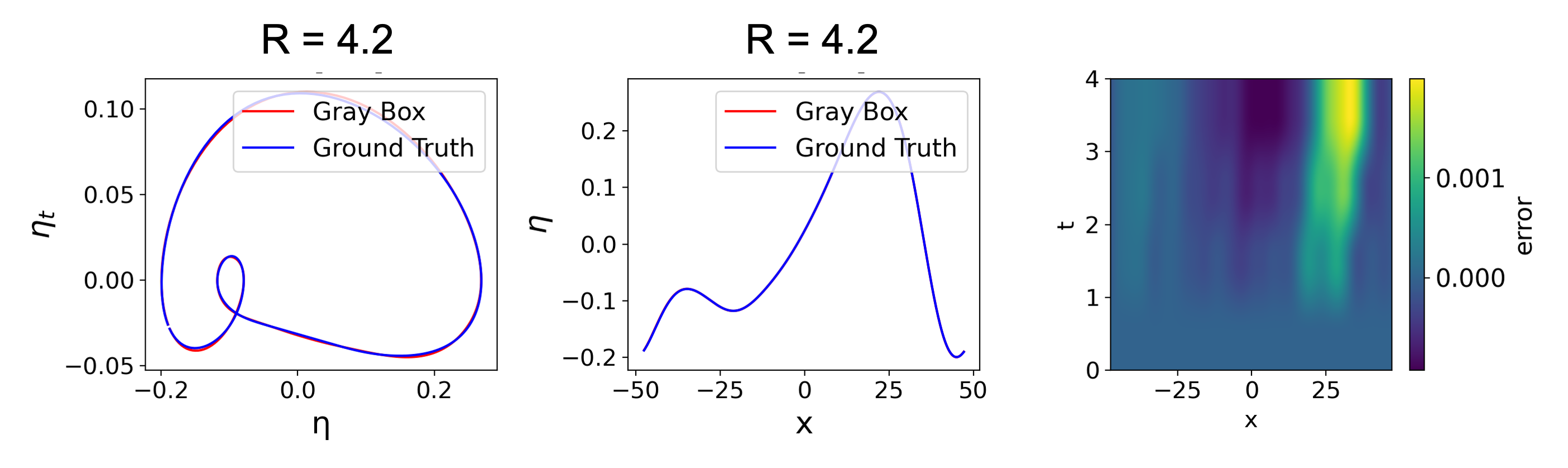}
\caption{R=4.2; Ground truth wave vs additive Gray box model attractor comparison (left); wave comparison in a specific time-step (center); Absolute error at each point in space and time between "Gray box I" model and ground truth.}
\label{fig:Gray}
\end{figure}

This approach maintains the physical insight already offered by the approximate equations, but improves its accuracy in a data-driven fashion. The predictions of this corrected model, referred to as a "Gray box" model, to contrast with the "Black box" model presented before, are visually very close to the ground truth as shown in Fig.~\ref{fig:Gray}, for $R=4.2$.

\subsection{Gray Box model II: a Functional Correction.}
Exploiting further the physical insight of the KS, even in parameter ranges where it is inaccurate, it is possible to use local, in space and/or in time, values of the KS right-hand-side to approximate the "correct" right-hand-side of the data-driven PDE. Now, a neural network is trained to predict the time derivative of the amplitude, given a few locally nearby values of the KS time-derivative, or a few of its derivatives with respect to the dependent variable, $\eta$, or a few nearby values of its spatial partial derivatives, e.g. $\eta_x$, $\eta_{xx}$. Several flavours of this approach are implemented:
\begin{equation}
\label{eq:functional}
  \begin{split}
  &  \left(\frac{\partial{\eta}}{\partial t}\right)^{NN}=f\left(f^{KS},\partial{f^{KS}}/\partial{\eta},\partial{f^{KS}}/\partial{\eta_{x}},R\right)\\
   & \left(\frac{\partial{\eta}}{\partial t}\right)^{NN}=f\left(\partial{f^{KS}}/\partial{\eta},\partial{f^{KS}}/\partial{\eta_x},\partial{f^{KS}}/\partial{\eta_{xx}},R\right)\\
   & \left(\frac{\partial{\eta}}{\partial t}\right)^{NN}=f\left(f^{KS}_{j},f^{KS}_{j-1},f^{KS}_{j+1},R\right)\\
   & \left(\frac{\partial{\eta}}{\partial t}\right)^{NN}=f\left(f^{KS}_{t},f^{KS}_{t-1},f^{KS}_{t-2},R\right)\\\\
  \end{split}
\end{equation}
\noindent
In the last two examples, the subscripts, $j$, $j-1$, $j+1$ signify points in space where the value of the the KS right-hand-side, $f^{KS}$ is taken in the same time-step; whereas the subscripts, $t$, $t-1$, $t-2$ stand for different nearby points in time, where the value is taken at the same point in space.

This idea is inspired by the Takens embedding theorem ~\cite{Takens1981,Whitney1936}, which allows us to recover missing variables and reconstruct the dynamics of a system,  given a short time history of the variables that can be measured. Here this short time history is replaced by values of $f^{KS}$ and some of its partial derivatives, e.g. $\partial f^{KS}/\partial{\eta}$ and $\partial f^{KS}/\partial{\eta_x}$. Alternatively, one can consider the value of $f^{KS}$ at the same point in time but in three different nearby spatial points, or in the same point is space but in three different nearby time-steps.

\begin{figure}[ht]
\centering
\includegraphics[width=\columnwidth]{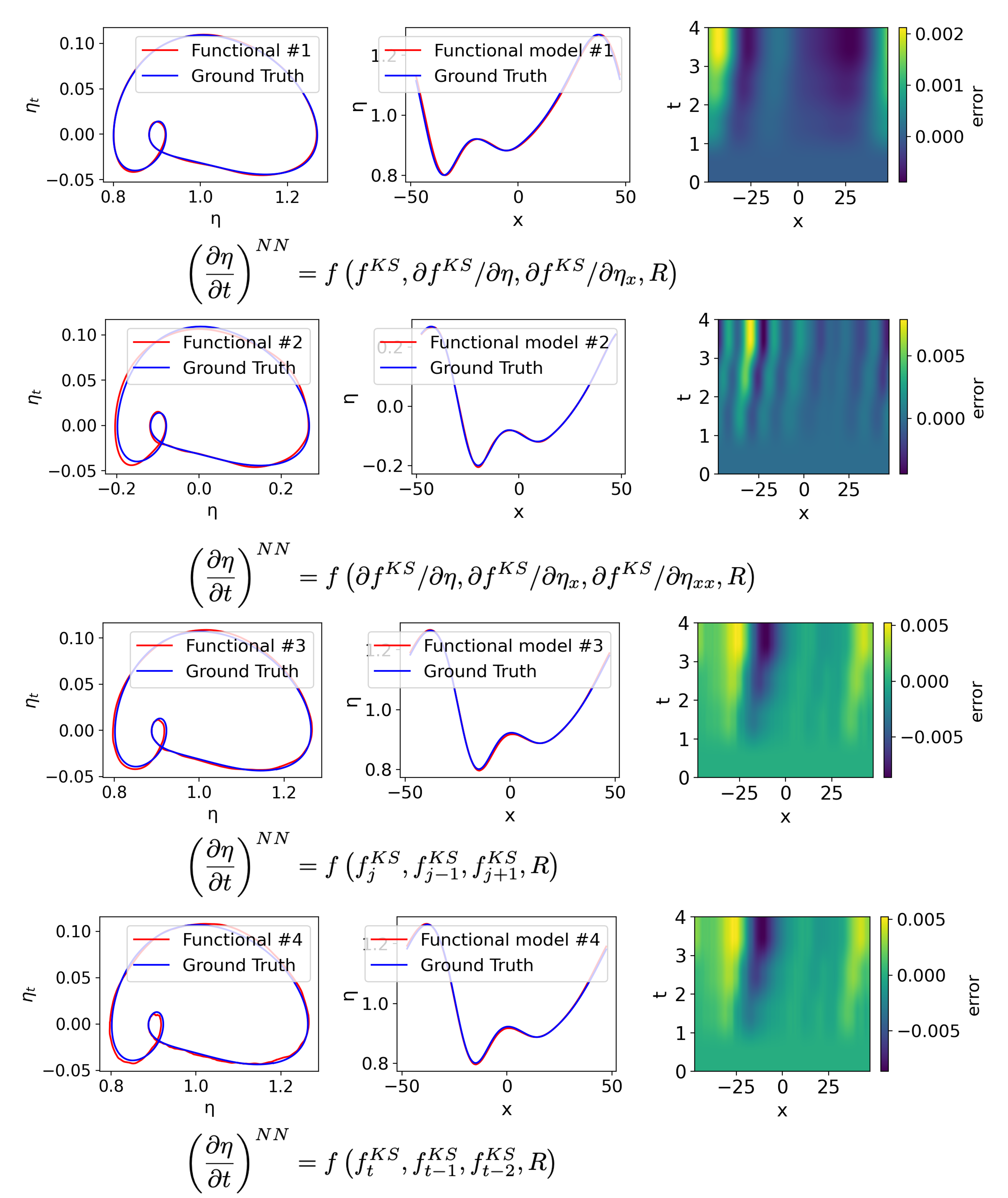}
\caption{R=4.2; For each case of Functional Gray box model  (denoted below each row of figures): Ground truth wave vs Functional Gray box model attractor comparison (left); wave comparison at a specific time-step (center); Absolute error at each point in space and time between "Gray box II" and ground truth.}
\label{fig:Func}
\end{figure}

The performance of the alternative functional models is shown in Fig.~\ref{fig:Func}, for the same parameter value, R=4.2. The error is slightly higher than in the previous case of the additive residual model.

\subsection{Out of sample predictions: A "Gappy DMAP" approach}
We now shift our focus to exploiting NS data to recover missing information. Missing data is a critical problem in applications in flow measurement and monitoring. For example, in film flow applications, it is often easy for an experimentalist to measure the film height, whereas being able to evaluate the detailed underlying flow field is a significantly more difficult task, if not impossible in the case of opaque liquids. Our goal here is to exploit NS data to derive a predictive tool, e.g. for the full velocity field or even the flow parameter, the Reynolds number, from only partial information. More importantly, to be able to do so as efficiently as possible and without having to care too much about the sensor positions.

The proposed approach is inspired by Gappy POD ~\cite{everson1995karhunen}, according to which it is possible to recover missing information from a vector that we know belongs to the subspace spanned by a few predetermined POD modes, by performing efficient interpolation in this reduced subspace. Here, the same concept is demonstrated, but also the notion of deriving a reduced description of the data with nonlinear manifold learning, in this case with Diffusion Maps ~\cite{r19,r20,r21} (details can be found in Methods). The added benefit is twofold: firstly, Diffusion Maps identify a parsimonious parametrization of the reduced subspace, which requires significantly less modes than POD, especially if the data belong to a curved manifold. The second benefit it related to the fact that in Gappy POD the accuracy of the method is critically influenced by the "location" of the known elements of the vector. The reason is purely numerical, and has to do with the condition number of the Gappy matrix $M = ($m$ \cdot \bf{\Phi})^{'} \cdot ($m$ \cdot \bf{\Phi})$, with $\bf{\Phi}$ the selected POD basis and $m$ the mask matrix that defines which elements of the vector are known.

\begin{figure}[ht]
\centering
\includegraphics[width=\columnwidth]{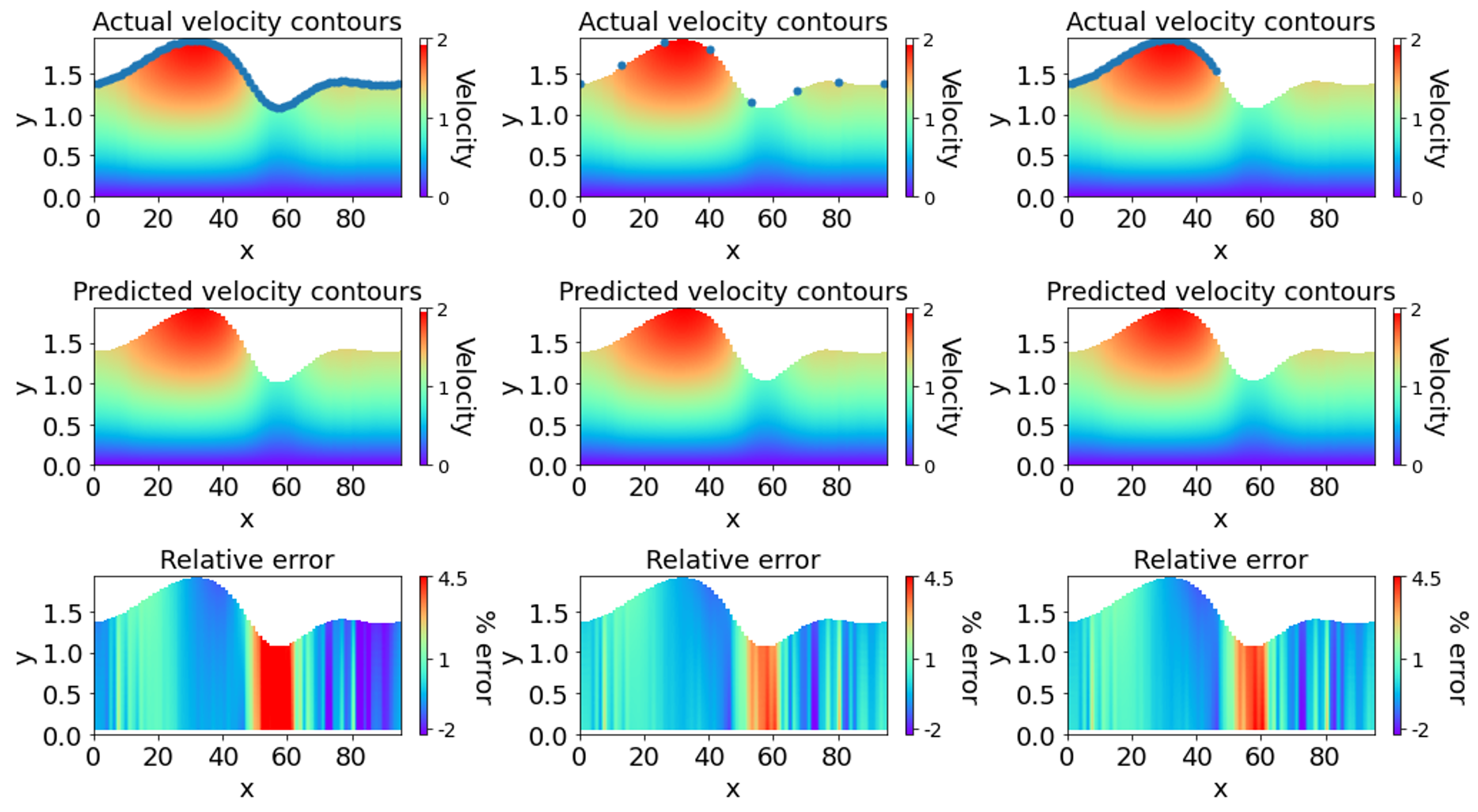}
\caption{Double DMAPS; R=3.5; Each column of figures presents the actual velocity contours (top), predicted velocity contours (center) and error(bottom). The blue dots in the top figures signify points where the value of the amplitude is considered known. On the left column, the value of the amplitude in 80 points is considered known and the maximum error is $4\%$; At the center, the value at 8 equidistant points is considered known and the maximum error is $4\%$; on the right, the value at 40 points in the first half of the wave are considered known and the maximum error is again close to $4\%$}
\label{fig:DoubleDMAPS}
\end{figure}

The implementation of "Gappy" Diffusion Maps, starts by identifying a parametrization of the manifold, where the data-set belongs to. It is found that three diffusion coordinates are enough to describe any vector in the data-set. Then, a second round of Diffusion Maps is implemented, in conjunction to Geometric Harmonics interpolation (details can be found in Methods), in order to map from any point on the reduced space to the high-dimensional ambient space. Having established the methods for mapping between the ambient and the reduced space, it is now possible, given partial information, to find first the corresponding reduced coordinates; and then the entire ambient vector, including the missing information. The accurate performance of this workflow is demonstrated in Fig.~\ref{fig:DoubleDMAPS}, where three cases are examined: (i) 80 points along the interface are known, from which the velocity and parameter value is recovered, with a maximum error of $4\%$; (ii) 8 points, evenly distributed along the interface are known, from which again the velocity values and the parameter is recovered with a maximum error of $4\%$; (iii) 40 points are known belonging to only half of the interface shape, leading to prediction of the velocities and the parameter with $4\%$ error.

\begin{figure}[h!]
\centering
\includegraphics[width=\columnwidth]{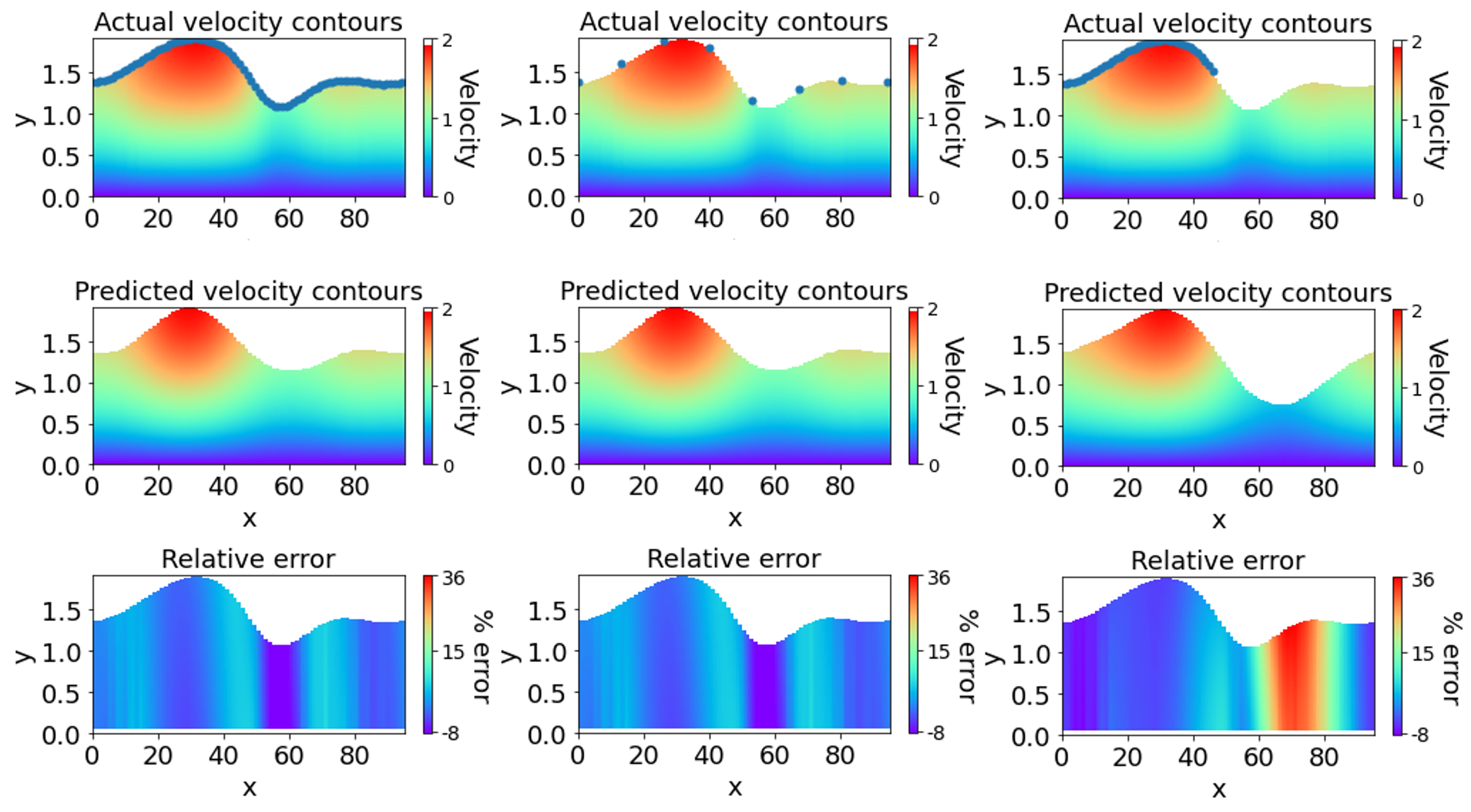}
\caption{Gappy POD; R=3.5; Each column of figures presents the actual velocity contours (top), predicted velocity contours (center) and error(bottom). The blue dots in the top figures signify points where the value of the amplitude is considered known. On the left column, the value of the amplitude in 80 points is considered known and the maximum error is $10\%$; at the center, the value at 8 equidistant points is considered known and the maximum error is $10\%$; on the right, the value at 40 points in the first half of the wave are considered known and the maximum error is close to $35\%$ }
\label{fig:GappyPOD}
\end{figure}

The same computational experiments are conducted with Gappy POD and the results are shown in Fig.~\ref{fig:GappyPOD}. First a POD basis is determined, based on the reconstruction error of the data-set, which leads to a basis with 6 POD vectors. The same points along the interface as before are considered known: (i) with 80 points along the interface, the maximum prediction error for the unknown velocity and parameter values is $10\%$;(ii) with 8 equidistant points along the interface, the maximum prediction error is approximately the same; (iii) when 40 points along half of the interface are considered, then the maximum error soars to $35\%$ and the predicted wave shape and velocity distribution is visibly inaccurate.

\section{Discussion}
In this work we presented three different strategies for deriving accurate and economical surrogates of the amplitude evolution of falling thin films. 

The first option is purely data-driven and physics-agnostic, and relies on learning, as a "Black box", an amplitude PDE from observed NS data over a range of R values. In essence, the right-hand-side of the PDE is substituted by an Artificial Neural Network, which can then be integrated in time, for various times, different initial conditions and parameter values. 

As an alternative, we propose using a low-fidelity model, here the KS equation, in order to infuse physical intuition into the learned model. This is achieved in two different ways: the first, the additive "Gray Box" approach, uses a few high-fidelity data (results of the NS) to calibrate the KS, by learning, by way of a neural network, an \textit{additive correction} rather that the entire right-hand-side.

The second strategy, the so-called "Functional Gray box" approach, is inspired by Takens' embedding theory, and proposes learning the \textit{accurate} amplitude dynamics from a few observations of an \textit{inaccurate} model right-hand side, such as the KS. Four different options are presented, utilizing either the RHS of the KS and some of its derivatives, or the value of the RHS operator at three points in space (at the same time instance) or in time (for the same point in space). This last approach is a demonstration of how a model that is qualitatively close to the ground truth, but quantitatively off, can be leveraged, in the data-driven era, into a more accurate and efficient learned model.

Finally, we presented a "Gappy DMAP" methodology, the nonlinear counterpart of Gappy POD, which allows us to infer quantities that are inaccessible to measuring devices, such as the velocity profile of the fluid below the surface, when only some measurements are known, such as the height of the thin film at certain points. This may be trivial for low R values, since the interface height is "slaved" to the velocity. Nevertheless, for slightly higher R values, the amplitude is no longer a function of just the position (hence surrogate models with more that one equation become necessary in this flow regime) ~\cite{Kalliadasis2012,oron2002nonlinear,shklyaev2017longwave, chang1986nonlinear}.
%%%YGK  references ?  - done

The benefits of the proposed approach are twofold: nonlinear manifold learning methods, such as DMAPs, yield a more parsimonious description of the manifold, requiring only a few coordinates to accurately reconstruct the original data. In contrast, we demonstrate that Gappy POD, requires higher-dimensional hyper-planes to span the data. 
The second advantage of Gappy DMAPs is related to the choice of known values: some consideration is necessary for choosing points that contain rich enough information in order to achieve accurate reconstruction; it is, nevertheless, less sensitive to the position of the provided measurements than its linear counterpart.
%%%YGK
%% Eleni, this is wrong - you choose bad points, the dmaps will have many harmonics before you can see new information - think about it and we discuss - SEE ABOVE PHRASE

\section{Methods}
\subsection{The Navier-Stokes equations for flow on an vertical plane }
The flow of a liquid on an inclined plane is described, in two dimensions, by conservation equations for mass and momentum, written in dimensionless form:
\begin{equation}
\label{eq:NavierStokes}
  \begin{split}
  &  \nabla \cdot \textbf{u} = 0, \\
   & R \frac{\partial \textbf{u}}{\partial t}= - R (\textbf{u} \cdot \nabla \textbf{u}) - \nabla P + \nabla^ 2\textbf{u} + F \textbf{g}, \\
  \end{split}
\end{equation}

\noindent
where $\textbf{u}=\left( u_x, u_y\right)^T$ and $P$ are the dimensionless velocity vector and pressure, respectively, and $\nabla=\left( \partial_x, \partial_y \right)$ denotes the gradient operator for Cartesian coordinates. We also define the unit gravity vector $\textbf{g}=\left(\sin\theta,-\cos\theta\right)^T$. Using the characteristic Nusselt scales for the velocity vector and all lengths
\begin{equation}
\label{eq:Scales}
U_N=\left(\frac{Q^{2} \rho g sin(\theta)}{3\mu}\right)^{1/3}, \\ 
H_N=\left(\frac{3\mu Q}{ \rho g sin(\theta)}\right)^{1/3}
\end{equation}

\noindent
the dimensionless groups that emerge are the Reynolds number $R$, the Weber number $W$, and the Stokes number $F$, defined as
\begin{equation}
\label{eq:Groups}
R = \frac{\rho Q}{\mu}, \; 
W = \frac{\sigma}{\rho g H_N^2}, \;
F = \frac{\rho g H_N^3}{\mu Q}.
\end{equation}

\noindent Here, $\rho$, $\mu$ and $\sigma$ are the liquid density, the viscosity and the liquid/air surface tension, respectively, while $Q$ denotes the volumetric flow rate per unit length normal to the cross section.

Along the liquid/air interface, a normal stress balance between capillary force and viscous stress is applied

\begin{equation}
   \textbf{n}\cdot \textbf{T} = W F \kappa \; \textbf{n}.
  \label{eq:interfacepressure}
\end{equation}
\noindent
Here, the total stress tensor is defined as $\textbf{T}=-$P$\textbf{I}+\left(\nabla \textbf{u}+\left(\nabla\textbf{u}\right)^T\right)$, \textbf{n} is the unit vector normal to the interface, outward with respect to the film. Note that in Eq. \ref{eq:interfacepressure} the ambient pressure has been set equal to zero (datum pressure) without loss of generality. The mean curvature is $\kappa=-\nabla_s \cdot \textbf{n}$ %$2H_m=-\nabla_s \cdot \textbf{n}$
, with $\nabla_s=\left(\textbf{I}-\textbf{nn}\right)\cdot \nabla$. 
The rest of the boundary conditions include the no slip condition at the liquid/solid plane interface (denoted as AB in Fig.~\ref{fig:film}) and periodic boundary conditions at the lateral domain boundaries (AD and BC in Fig.~\ref{fig:film}).  

The kinematic boundary condition, which specifies that the velocity of the interface that is normal to the boundary is equal to the velocity of the fluid that is normal to the boundary, ensuring no mass transfer through the interface, completes the set of governing equations:

\begin{equation}
   \textbf{n}\cdot (h_t \textbf{e}_y)=\textbf{n}\cdot \textbf{u}
  \label{eq:kinematic}
\end{equation}

%\begin{equation}
%   \textbf{n}\cdot h_t=\textbf{n}\cdot \textbf{u}
%  \label{eq:kinematic}
%\end{equation}

\subsection{Transformations between Navier-Stokes and Kuramoto-Sivashinsky scales}

The $(x,t)$ NS frame of reference is mapped on the KS $(\xi,\tau)$ through the following expressions

\begin{equation}
    \xi=\epsilon(x-Ft)\\ \quad \mbox{and} \quad \tau = \frac{\epsilon^4 W F}{12} t, 
\end{equation}
while the interfacial height, $h(x,t)$ is related to the amplitude $\phi(\xi,\tau)$ as 
\begin{equation}
    h(x,t) \approx 1+\epsilon\frac{RF}{15}\phi(\xi,\tau).\\
\end{equation}

To be able to compare the results between NS and KS we employ the chain rule and derive the appropriate transformations for the time and spatial derivatives. 
To transform NS data to the KS formulation the following expressions can be used: 
\begin{equation}
  \begin{split}
   \frac{\partial^{\beta} \phi}{\partial \xi^{\beta}}&=\frac{1}{\epsilon^{\beta+1}}\frac{15}{RF}\frac{\partial^{\beta} h}{\partial x^{\beta}},\\
    \frac{\partial \phi}{\partial \tau}&=\frac{180}{\epsilon^5 W R F^2}\frac{\partial h}{\partial t}+\frac{180}{\epsilon^5 W R F}\frac{\partial h}{\partial x}.
   \end{split}
\end{equation}
Inversely, to map KS data to the NS formulation the following expressions can be used:
\begin{equation}
   \begin{split}
    \frac{\partial^{\beta} h}{\partial x^{\beta}}&=\epsilon^{\beta+1}\frac{ R F}{15}\frac{\partial^{\beta} \phi}{\partial \xi^{\beta}},\\
    \frac{\partial h}{\partial t}&=-\frac{\epsilon^2 R F^2}{15}\frac{\partial \phi}{\partial \xi}+\frac{\epsilon^5 W R F^2}{180}\frac{\partial \phi}{\partial \tau}.\\
   \end{split}
\end{equation}

\subsection{Derivation of the KS from the NS}
The detailed derivation of KS from NS can be found in ~\cite{shlang1982irregular,chang1986nonlinear,brown1992} and it is summed up here for completeness. It is based in the following assumptions:

\begin{itemize}
    \item The film is thin, $\epsilon=2 \pi H_N/L \ll 1$
    \item The waves are long, $\frac{\partial h}{\partial x} \ll 1$
    \item The mean height of the film is much larger than the deviation from the mean 
\end{itemize}

Under these assumptions, it is possible to exploit the small parameter $\epsilon$ and employ a perturbation expansion for all dependent variables, i.e. velocities, pressure and interfacial height; e.g. the interfacial height is given by $h\approx 1+\epsilon \eta + O(\epsilon^2)$, where $\eta$  denotes the deviation from the mean film height. Restricting to the case of laminar flow with $R=O(1)$, $F=O(1)$ and $W=O(\epsilon^{-2})$ and neglecting higher order terms, the kinematic condition (written in terms of the deviation, $\eta$) may be reduced
% It is assumed that \gk{$\epsilon=2 \pi H_N/L$}, the wavenumber, is small (long wave assumption) and higher order terms are eliminated from the variables. As a result, it holds that: $h\approx 1+\epsilon \eta$. 
%  The kinematic condition is written as a function of $\eta = \frac{h-1}{\epsilon}$, and the flow is restricted to laminar flow regimes where: $Re=O(1)$, $F=O(1)$, and $\epsilon^2W=O(1)$, as a result, the kinematic condition is simplified 
to: $\eta_t+F\eta_x=0$, indicating that waves travel with speed $-F$. Taking this into account, a new variable $\xi=\epsilon (x-Ft)$ can be introduced to obtain constant shape waves traveling with speed $F$. Moreover, the amplitude is rescaled according to $\phi=\frac{15}{RF}\eta$. Finally, since it is known that the wave amplitude vary on a slow time scale compared with the traveling motion, a change in the time variable is introduced, i.e. $\tau = \frac{\epsilon^4 W F}{12} t$.
In  the end, the KS (Eq.~\ref{eq:KS}) is obtained as a function of the new defined variables.

\subsection{Diffusion Maps}
\label{sec:Diffusion_Maps}
Diffusion maps ~\cite{r19,r20,r21} is a framework that can (based upon diffusion processes) facilitate discovering meaningful low-dimensional intrinsic geometric descriptions of data sets, even when the data is high-dimensional, nonlinear and/or corrupted by (relatively small) noise. The method is based on the construction of a Markov transition probability matrix, corresponding to a random walk, on a graph whose vertices are the data points, with transition probabilities being the local similarities between pairs of data points. The leading few eigenvectors of the sparse 
%%%YGK  why is the Markov matrix sparse ? 
Markov matrix can be used as data-driven coordinates that provide a reparametrization of the data. %that is called diffusion maps (DMAPS henceforth in the text). 

To construct a low-dimensional embedding for a data set $\mathbf{X}$ of M individual points (represented as d-dimensional real vectors \(x_1,..., x_M\)), a similarity measure \(d_{ij}\) between each pair of vectors \(x_i, x_j\) is computed. The standard Euclidean distance or the Euclidean norm may be considered to this end. By using this similarity measure,% a pairwise affinity function \(w = w (d_{ij})\) is computed with \(w(0)=1\) and \(w(d)>0\) and monotonically decreasing for \(d>0\). 
an affinity matrix is constructed. A popular choice is the Gaussian kernel
\begin{equation*}
w(i,j)=exp\left[-\left(\frac{||x_i-x_j||}{\epsilon_{\delta}}\right)^2\ \right]
\end{equation*} where $\epsilon_{\delta}$ defines a  scale hyperparameter which quantifies the local similarity for each data point.   %characteristic scale which quantifies the “locality” of the neighborhood within which Euclidean distance can be used as the basis of a meaningful similarity measure.
To recover a parametrization regardless of the sampling density, the normalization
\[\widetilde{\textbf{W}} = \textbf{P}^{-\alpha} \textbf{W} \textbf{P}^{-\alpha}\]
is performed, where $P_{ii} = \sum_{j=1}^M W_{ij}$ and $\alpha=1$ to factor out the density effects.
A second normalization applied on $\widetilde{\textbf{W}}$, %
\[\textbf{K}=\textbf{D}^{-1} \widetilde{\textbf{W}}\] 
gives a \(M\times M\) Markov matrix \textbf{K}; where \(\textbf{D}\) is a diagonal matrix, collecting the row sums of matrix \(\widetilde{\textbf{W}}\).
%approach, a \(M\times M\) Markov matrix \textbf{K} is constructed using the symmetric matrix \(W_{ij}={w(x_i, x_j)}\) as \[\textbf{K}=\textbf{D}^{-1} \textbf{W}\] where \(\textbf{D}\) is a diagonal, collecting the row sums of matrix \(\textbf{W}\).
%
The stochastic matrix \(\textbf{K}\) has a set of real eigenvalues \hbox{\(1=\lambda_1\geq\ ... \geq\lambda_M\)} with corresponding eigenvectors $\phi_i$. % The M-dimensional representation of a particular n-dimensional data vector \(x_i\),  is given by the diffusion map: \[\Psi_M(x_i)=[\lambda_1^t\phi_{i,1},\lambda_2^t\phi_{i,2},...,\lambda_M^t\phi_{i,M}]\] as it assigns a vector of M coordinates to each data point $x_i$ through a mapping which is only defined on the M recorded data vectors. 

To check if model (variable) reduction can be achieved, the number of retained eigenvectors has to be appropriately truncated. In practice, it is useful to consider that not all obtained eigenvectors parametrize independent directions, but rather most of them can be considered as spanning the same directions with different frequencies. Eigenvectors that parametrize the same directions in this context are called \textit{harmonics} and the ones that parametrize independent directions \textit{non-harmonics}. %of merely a few, thus not proving a new direction on the dataset.
A minimal representation of the DMAP space is made possible by carefully selecting the non-harmonic coordinates, which do not necessarily correspond to the most dominant eigenmodes of the Markov matrix. This is a stark difference between Diffusion Maps and its linear counterpart, Proper Orthogonal Decomposition or Principal Component analysis, where the dominant modes are retained for the truncated representation of the data. If the number of the \textit{non-harmonic} eigenvectors is less than the number of the ambient space dimensions then model (variable) reduction is achieved.

A proposed algorithm for identifying the non-harmonic eigenvectors is presented in \cite{r25}, based on local linear regression. In a nutshell, a local linear function is used in order to fit the DMAP coordinate $\phi_k$ as a function, \(f\), of the previous vectors \(\widetilde{\Phi}_{k-1}= [\phi_1,\phi_2,...,\phi_{k-1}]\). If $\phi_k$ can be accurately expressed as function of the other DMAP coordinates, then it does not represent a new direction on the dataset, and is omitted for dimensionality reduction. On the contrary if $\phi_k$ cannot be expressed as a function of the previous eigenvectors then $\phi_k$ is a new independent eigendirection that must be retained for a parsimonious representation of the data. To quantify the accuracy of the fit, the following metric is used:\[r_k=\sqrt{\frac{\sum_{i=1}^{n}(\phi_k(i)-f(\widetilde{\Phi}_{k-1}(i)))^2}{\sum_{i=1}^{n}(\phi_k(i))^2)}}\]
A small value of \(r_k\) is associated with a  $\phi_k$ that is a harmonic function of the previous eigenmodes, whereas a higher value of $r_k$ signifies that $\phi_k$ is a new independent direction on the data manifold. It has been shown in \cite{r25} that selecting only the eigenvectors that correspond to higher values of $r_k$ leads to a parsimonious representation of the data. Eventually, the vector $x_i$ is mapped to a vector whose first component is the \textit{i}-th component of the first selected nontrivial eigenvector, whose second component is the \textit{i}-th component of the second selected nontrivial eigenvector, etc. 

To map a new point, $x_{new}$, from the ambient space to DMAP space, a mathematically elegant approach known as the Nyström extension, introduced in [3] is used, summarized here for completeness. The starting point of the Nyström extension is to compute the distances, $d({\cdot, x_{new}})$,  between the new point, $x_{new}$, and the $M$ data points in the original data set, the same normalizations used for DMAP need to be applied also here. The Nystr\"om extension formula reads

\[\phi_j(x_{new})=\lambda_j^{-1}\sum_{i=1}^{M}\tilde{k}(x_i,x_{new})\phi_{j}(x_{i}),\]
where $\lambda_j$ is the \textit{j}-th eigenvector and $\phi_j(x_i)$ is the \textit{i}-th component of the j-th eigenvector.  

% \[w(i,new)=\exp{\left[-\left(\frac{d_{i,new}}{\epsilon}\right)^2\right]}\]
% Setting $\tilde{k}(i,new)=\left(\sum_{k=1}^{M}w(k,new)\right)^{-1}w(i,new)$, the jth DMAP coordinate of the new vector $y_{new}$ is given by:\[\phi_j(y_{new})=\lambda_j^{-1}\sum_{i=1}^{m}\tilde{k}(i,new)\phi_{j}(x_{i})\] Clearly, this extension procedure will return a result even if $y_{new}$ is not chosen to be exactly on our low-dimensional manifold.  \textcolor{red}{My suggestion is: We might need to elaborate more on this or avoid saying it.}

\subsection{Geometric Harmonics}
Geometric Harmonics was introduced in \cite{r19}, inspired by the Nyström Extension as a scheme for \textit{extending} functions defined on data $\mathbf{X}$, $f(\mathbf{X}):\mathbf{X} \to \mathbb{R}$, for $x_{new} \notin \mathbf{X}$. This out-of-sample extension is achieved by using a particular set of basis functions called Geometric Harmonics. Those functions are computed as eigenvectors  of the symmetric \(M\times M\) \(  \textbf{W}\) matrix.  The eigendecomposition of the symmetric and positive semidefinite matrix ${\mathbf{W}}$ leads to a set of orthonormal eigenvectors $\mathbf{\psi}_1,\mathbf{\psi}_2, \dots, \mathbf{\psi}_M$ with non negative eigenvalues $\sigma_1 \geq \sigma_2 \geq \dots \geq \sigma_M \geq 0 $. 

From this set of eigenvectors, to avoid numerical issues, we consider a truncated subset  \hbox{\(S_\delta\) = (\(\alpha\) :  \({\sigma_{\alpha}\geq\delta\sigma_1}\))} where \(\delta>0\). The extension of $f$ for a new point $x_{new}$
%is defined, the extension of a function, \(f\), defined only at some sample points in \(Z\subset X\) to an arbitrary new point in \(Z\), is 
is accomplished by first projecting the function of interest in the (truncated) computed set of eigenvectors \[f\rightarrow P_\delta f =\sum_{\alpha\in S_\delta}\langle f,\psi_\alpha \rangle \psi_\alpha\] and then extending the function $f$ for $x_{new} \notin \mathbf{X}$

\begin{equation*}
(Ef)(x_{new})= \sum_{\alpha \in S_{\delta}}\langle f,\psi_{\alpha} \rangle\Psi_\alpha(x_{new})
%  \quad\text{where}\quad 
% \Psi_\alpha(x_{new})=\lambda_\alpha^{-1}\sum_{i=1}^{M}w(x_{new},x_i)\psi_{\alpha}(x_{i}) %,\alpha \in S_\delta
\end{equation*}
%and w(y_i,y)=e^{-\left(\frac{d_i}{\epsilon}\right)^2},d_i=||y_i-y||
%where \[\Phi_\alpha=\lambda_\alpha^{-1}\sum_{i=1}^{M}w(y_i,y)\phi_{i,\alpha},\alpha \in S_\delta \end{equation}
where
\begin{equation*}
\Psi_\alpha(x_{new})=\lambda_\alpha^{-1}\sum_{i=1}^{M}w(x_{new},x_i)\psi_{\alpha}(x_{i}) %,\alpha \in S_\delta
\end{equation*}
and \[w(x_{new},x_i)=exp\bigg[{-\left(\frac{d_i}{\tilde{\epsilon}}\right)^2\bigg]},d_i=||x_{new} - x_i||_2\]

\subsection{Double Diffusion Maps and their Latent Harmonics}
A \textit{slight} twist of the  Geometric Harmonics is presented in this section. As discussed above, Geometric Harmonics constructs an input-output mapping between the ambient coordinates $\mathbf{X}$ and a function of interest $f$ defined on $\mathbf{X}$. However, it is possible if the data are lower dimensional, to construct a map in terms of only the non-harmonic eigenvectors. This is achieved similar to the \textit{traditional} Geometric Harmonics, by firstly constructing an affinity matrix

\begin{equation*}
\overline{w}(i,j)=exp\left[-\left(\frac{||\phi_i-\phi_j||}{\epsilon^\star}\right)^2\ \right].
\end{equation*}
In this case the affinity matrix is constructed in terms of only the non-harmonic DMAPs coordinates. To distinguish the notation between Geometric Harmonics and Double Diffusion Maps we will use $\overline{\cdot}$.
%%%YGK is it dot or bar ?
As in the \textit{traditional} Geometric Harmonics the function $f$ is projected to a truncated set of the obtained eigenvectors

\[f\rightarrow P_\delta f =\sum_{\beta\in \overline{S}_\delta}\langle f,\overline{\psi}_\beta \rangle \overline{\psi}_\beta.\] 
The extension of $f$ for $\mathbf{\phi}_{new}$ is achieved by firstly extending the values of the Geometric Harmonic functions $\Psi_{\beta}$ for $\mathbf{\phi}_{new},$

\begin{equation*}
\overline{\Psi}_\beta(\mathbf{\phi}_{new})=\overline{\sigma}_\beta^{-1}\sum_{i=1}^{M}\overline{w}(\phi_{new},\phi_i)\overline{\psi}_{\beta}(\mathbf{\phi}_{i}), %,\alpha \in S_\delta
\end{equation*}
and then estimating the value of $f$ at $\mathbf{\phi}_{new}$
\begin{equation*}
(Ef)(\mathbf{\phi}_{new})= \sum_{\beta \in \overline{S}_{\delta}}\langle f,\overline{\psi}_{\beta} \rangle\overline{\Psi}_\beta(\mathbf{\phi}_{new})
\end{equation*}

\subsection{Gappy POD}
In this section the Gappy POD method is summarized for completeness. Consider a data set $\mathbf{X}$ of M vectors (represented as d-dimensional real vectors \(x_1,..., x_M\)).
A POD basis, $\mathbf{\Phi} \in \Re^{N \times M}$, of $\mathbf{X}$ is computed, such that $\mathbf{X}$ can be approximated as a linear combination of p vectors:

$\widetilde{\mathbf{X}}= \sum_{j=1}^{p} c^j\mathbf{\Phi}^j$ 

or, in matrix-vector format:

$\widetilde{\mathbf{X}}=\bf{\Phi}\cdot c$

The size of the truncated POD basis $\mathbf{\Phi}$ is selected based on the error between the actual  vector $\bf{X}$ and the reconstructed approximation $\tilde{\bf{X}}$ :

reconstruction error = $\lVert  \bf{X}-\tilde{\bf{X}}  \rVert$

Consider now a vector $X^{'}$ that is spanned by the same basis $\Phi$ and that only m values of this vector are known, so that the partial vector $X_{partial}^{'}$ can be defined: 

$X_{partial}^{'}=$m$\cdot X^{'},$m$ \in \Re^{m\times N}$  

The goal is to find coefficients $c^{'}$, such that an approximation $\tilde{X^{'}}$ of the vector $X^{'}$ can be defined as:

$\tilde{X^{'}}=X^{'} \cdot c^{'}$;

then 

$X_{partial}^{'} \approx  $m$ \cdot X^{'} \cdot c^{'}$.
\noindent 
Finding the values of $c^{'}$ that satisfy the above leads to an optimization problem solved through the linear system:

$M \cdot c^{'} = ($m$ \cdot  \bf{\Phi})^{'} \cdot $X$_{partial}^{'},$

with $M = ($m$ \cdot  \bf{\Phi})^{'} \cdot ($m$ \cdot  \bf{\Phi})$.

\subsection*{Acknowledgment}
This work was partially supported by the US AFOSR and by the US DOE (IGK). CML received the support of a "la Caixa" Foundation Fellowship (ID 100010434), code LCF/BQ/AA19/11720048.  EDK received funding from the European Union’s Horizon 2020 research and innovation programme under the Marie Skłodowska-Curie grant agreement No 890676  - DataProMat.

\subsection*{Author Contributions}
C.M.L. performed computational analyses, interpreted the data and wrote the manuscript,Y.P. 
performed computational analyses and wrote the manuscript, G.K developed 
computational fluid dynamics model, contributed to mathematical analyses 
and wrote the manuscript, E.D.K. performed computational analyses, 
interpreted the data and wrote the manuscript and I.G.K. conceived the 
research, supervised the study and wrote the manuscript.

\bibliography{sn-bibliography.bib}%


\begin{thebibliography}{43}
\ifx \bisbn   \undefined \def \bisbn  #1{ISBN #1}\fi
\ifx \binits  \undefined \def \binits#1{#1}\fi
\ifx \bauthor  \undefined \def \bauthor#1{#1}\fi
\ifx \batitle  \undefined \def \batitle#1{#1}\fi
\ifx \bjtitle  \undefined \def \bjtitle#1{#1}\fi
\ifx \bvolume  \undefined \def \bvolume#1{\textbf{#1}}\fi
\ifx \byear  \undefined \def \byear#1{#1}\fi
\ifx \bissue  \undefined \def \bissue#1{#1}\fi
\ifx \bfpage  \undefined \def \bfpage#1{#1}\fi
\ifx \blpage  \undefined \def \blpage #1{#1}\fi
\ifx \burl  \undefined \def \burl#1{\textsf{#1}}\fi
\ifx \doiurl  \undefined \def \doiurl#1{\url{https://doi.org/#1}}\fi
\ifx \betal  \undefined \def \betal{\textit{et al.}}\fi
\ifx \binstitute  \undefined \def \binstitute#1{#1}\fi
\ifx \binstitutionaled  \undefined \def \binstitutionaled#1{#1}\fi
\ifx \bctitle  \undefined \def \bctitle#1{#1}\fi
\ifx \beditor  \undefined \def \beditor#1{#1}\fi
\ifx \bpublisher  \undefined \def \bpublisher#1{#1}\fi
\ifx \bbtitle  \undefined \def \bbtitle#1{#1}\fi
\ifx \bedition  \undefined \def \bedition#1{#1}\fi
\ifx \bseriesno  \undefined \def \bseriesno#1{#1}\fi
\ifx \blocation  \undefined \def \blocation#1{#1}\fi
\ifx \bsertitle  \undefined \def \bsertitle#1{#1}\fi
\ifx \bsnm \undefined \def \bsnm#1{#1}\fi
\ifx \bsuffix \undefined \def \bsuffix#1{#1}\fi
\ifx \bparticle \undefined \def \bparticle#1{#1}\fi
\ifx \barticle \undefined \def \barticle#1{#1}\fi
\bibcommenthead
\ifx \bconfdate \undefined \def \bconfdate #1{#1}\fi
\ifx \botherref \undefined \def \botherref #1{#1}\fi
\ifx \url \undefined \def \url#1{\textsf{#1}}\fi
\ifx \bchapter \undefined \def \bchapter#1{#1}\fi
\ifx \bbook \undefined \def \bbook#1{#1}\fi
\ifx \bcomment \undefined \def \bcomment#1{#1}\fi
\ifx \oauthor \undefined \def \oauthor#1{#1}\fi
\ifx \citeauthoryear \undefined \def \citeauthoryear#1{#1}\fi
\ifx \endbibitem  \undefined \def \endbibitem {}\fi
\ifx \bconflocation  \undefined \def \bconflocation#1{#1}\fi
\ifx \arxivurl  \undefined \def \arxivurl#1{\textsf{#1}}\fi
\csname PreBibitemsHook\endcsname

%%% 1
\bibitem{glasser1997fully}
\begin{barticle}
\bauthor{\bsnm{Glasser}, \binits{B.}},
\bauthor{\bsnm{Kevrekidis}, \binits{I.}},
\bauthor{\bsnm{Sundaresan}, \binits{S.}}:
\batitle{Fully developed travelling wave solutions and bubble formation in
  fluidized beds}.
\bjtitle{Journal of Fluid Mechanics}
\bvolume{334},
\bfpage{157}--\blpage{188}
(\byear{1997})
\end{barticle}
\endbibitem

%%% 2
\bibitem{Kalliadasis2012}
\begin{bbook}
\bauthor{\bsnm{Kalliadasis}, \binits{S.}},
\bauthor{\bsnm{Ruyer-Quil}, \binits{C.}},
\bauthor{\bsnm{Scheid}, \binits{B.}},
\bauthor{\bsnm{Velarde}, \binits{M.G.}}:
\bbtitle{Falling {Liquid} {Films}}.
\bsertitle{Applied {Mathematical} {Sciences}},
vol. \bseriesno{176}.
\bpublisher{Springer},
\blocation{London}
(\byear{2012}).
\burl{http://link.springer.com/10.1007/978-1-84882-367-9}
\end{bbook}
\endbibitem

%%% 3
\bibitem{Pettas2019a}
\begin{botherref}
\oauthor{\bsnm{Pettas}, \binits{D.}},
\oauthor{\bsnm{Karapetsas}, \binits{G.}},
\oauthor{\bsnm{Dimakopoulos}, \binits{Y.}},
\oauthor{\bsnm{Tsamopoulos}, \binits{J.}}:
Viscoelastic film flows over an inclined substrate with sinusoidal topography.
  {I}. {Steady} state.
Physical Review Fluids
\textbf{4}(8)
(2019).
\doiurl{10.1103/PhysRevFluids.4.083303}
\end{botherref}
\endbibitem

%%% 4
\bibitem{Pettas2019b}
\begin{botherref}
\oauthor{\bsnm{Pettas}, \binits{D.}},
\oauthor{\bsnm{Karapetsas}, \binits{G.}},
\oauthor{\bsnm{Dimakopoulos}, \binits{Y.}},
\oauthor{\bsnm{Tsamopoulos}, \binits{J.}}:
Viscoelastic film flows over an inclined substrate with sinusoidal topography.
  {II}. {Linear} stability analysis.
Physical Review Fluids
\textbf{4}(8)
(2019).
\doiurl{10.1103/PhysRevFluids.4.083304}
\end{botherref}
\endbibitem

%%% 5
\bibitem{gonzalez-garcia98_ident_distr_param_system}
\begin{barticle}
\bauthor{\bsnm{Gonz{\'a}lez-Garc{\'i}a}, \binits{R.}},
\bauthor{\bsnm{Rico-Mart{\'i}nez}, \binits{R.}},
\bauthor{\bsnm{Kevrekidis}, \binits{I.G.}}:
\batitle{Identification of distributed parameter systems: a neural net based
  approach}.
\bjtitle{Computers \& Chemical Engineering}
\bvolume{22},
\bfpage{965}--\blpage{968}
(\byear{1998}).
\doiurl{10.1016/s0098-1354(98)00191-4}
\end{barticle}
\endbibitem

%%% 6
\bibitem{krischer93_model_ident_spatiot_varyin_catal_react}
\begin{barticle}
\bauthor{\bsnm{Krischer}, \binits{K.}},
\bauthor{\bsnm{Rico-Mart{\'i}nez}, \binits{R.}},
\bauthor{\bsnm{Kevrekidis}, \binits{I.G.}},
\bauthor{\bsnm{Rotermund}, \binits{H.H.}},
\bauthor{\bsnm{Ertl}, \binits{G.}},
\bauthor{\bsnm{Hudson}, \binits{J.L.}}:
\batitle{Model identification of a spatiotemporally varying catalytic
  reaction}.
\bjtitle{AIChE Journal}
\bvolume{39}(\bissue{1}),
\bfpage{89}--\blpage{98}
(\byear{1993}).
\doiurl{10.1002/aic.690390110}
\end{barticle}
\endbibitem

%%% 7
\bibitem{rico-martinez92_discr_vs}
\begin{barticle}
\bauthor{\bsnm{Rico-Mart{\'i}nez}, \binits{R.}},
\bauthor{\bsnm{Krischer}, \binits{K.}},
\bauthor{\bsnm{Kevrekidis}, \binits{I.G.}},
\bauthor{\bsnm{Kube}, \binits{M.C.}},
\bauthor{\bsnm{Hudson}, \binits{J.L.}}:
\batitle{Discrete- vs. continuous-time nonlinear signal processing of cu
  electrodissolution data}.
\bjtitle{Chemical Engineering Communications}
\bvolume{118}(\bissue{1}),
\bfpage{25}--\blpage{48}
(\byear{1992}).
\doiurl{10.1080/00986449208936084}
\end{barticle}
\endbibitem

%%% 8
\bibitem{kemeth22_learn_emerg_partial_differ_equat}
\begin{barticle}
\bauthor{\bsnm{Kemeth}, \binits{F.P.}},
\bauthor{\bsnm{Bertalan}, \binits{T.}},
\bauthor{\bsnm{Thiem}, \binits{T.}},
\bauthor{\bsnm{Dietrich}, \binits{F.}},
\bauthor{\bsnm{Moon}, \binits{S.J.}},
\bauthor{\bsnm{Laing}, \binits{C.R.}},
\bauthor{\bsnm{Kevrekidis}, \binits{I.G.}}:
\batitle{Learning emergent partial differential equations in a learned emergent
  space}.
\bjtitle{Nature Communications}
\bvolume{13}(\bissue{1}),
\bfpage{3318}
(\byear{2022}).
\doiurl{10.1038/s41467-022-30628-6}
\end{barticle}
\endbibitem

%%% 9
\bibitem{brunton16_discov_gover_equat_from_data}
\begin{barticle}
\bauthor{\bsnm{Brunton}, \binits{S.L.}},
\bauthor{\bsnm{Proctor}, \binits{J.L.}},
\bauthor{\bsnm{Kutz}, \binits{J.N.}}:
\batitle{Discovering governing equations from data by sparse identification of
  nonlinear dynamical systems}.
\bjtitle{Proceedings of the National Academy of Sciences}
\bvolume{113}(\bissue{15}),
\bfpage{3932}--\blpage{3937}
(\byear{2016}).
\doiurl{10.1073/pnas.1517384113}
\end{barticle}
\endbibitem

%%% 10
\bibitem{rudy17_data_driven_discov_partial_differ_equat}
\begin{barticle}
\bauthor{\bsnm{Rudy}, \binits{S.H.}},
\bauthor{\bsnm{Brunton}, \binits{S.L.}},
\bauthor{\bsnm{Proctor}, \binits{J.L.}},
\bauthor{\bsnm{Kutz}, \binits{J.N.}}:
\batitle{Data-driven discovery of partial differential equations}.
\bjtitle{Science Advances}
\bvolume{3}(\bissue{4}),
\bfpage{1602614}
(\byear{2017}).
\doiurl{10.1126/sciadv.1602614}
\end{barticle}
\endbibitem

%%% 11
\bibitem{long17_pde_net}
\begin{bchapter}
\bauthor{\bsnm{Long}, \binits{Z.}},
\bauthor{\bsnm{Lu}, \binits{Y.}},
\bauthor{\bsnm{Ma}, \binits{X.}},
\bauthor{\bsnm{Dong}, \binits{B.}}:
\bctitle{{PDE}-net: Learning {PDE}s from data}.
In: \beditor{\bsnm{Dy}, \binits{J.}},
\beditor{\bsnm{Krause}, \binits{A.}} (eds.)
\bbtitle{Proceedings of the 35th International Conference on Machine Learning}.
\bsertitle{Proceedings of Machine Learning Research},
vol. \bseriesno{80},
pp. \bfpage{3208}--\blpage{3216}.
\bpublisher{PMLR},
\blocation{Stockholmsmässan, Stockholm Sweden}
(\byear{2018}).
\burl{http://proceedings.mlr.press/v80/long18a.html}
\end{bchapter}
\endbibitem

%%% 12
\bibitem{raissi19_physic_infor_neural_networ}
\begin{barticle}
\bauthor{\bsnm{Raissi}, \binits{M.}},
\bauthor{\bsnm{Perdikaris}, \binits{P.}},
\bauthor{\bsnm{Karniadakis}, \binits{G.E.}}:
\batitle{Physics-informed neural networks: a deep learning framework for
  solving forward and inverse problems involving nonlinear partial differential
  equations}.
\bjtitle{Journal of Computational Physics}
\bvolume{378},
\bfpage{686}--\blpage{707}
(\byear{2019}).
\doiurl{10.1016/j.jcp.2018.10.045}
\end{barticle}
\endbibitem

%%% 13
\bibitem{Chen2018}
\begin{botherref}
\oauthor{\bsnm{Chen}, \binits{R.T.Q.}},
\oauthor{\bsnm{Rubanova}, \binits{Y.}},
\oauthor{\bsnm{Bettencourt}, \binits{J.}},
\oauthor{\bsnm{Duvenaud}, \binits{D.}}:
Neural Ordinary Differential Equations.
arXiv
(2018).
\doiurl{10.48550/ARXIV.1806.07366}.
\url{https://arxiv.org/abs/1806.07366}
\end{botherref}
\endbibitem

%%% 14
\bibitem{Vlachas2018}
\begin{barticle}
\bauthor{\bsnm{Vlachas}, \binits{P.R.}},
\bauthor{\bsnm{Byeon}, \binits{W.}},
\bauthor{\bsnm{Wan}, \binits{Z.Y.}},
\bauthor{\bsnm{Sapsis}, \binits{T.P.}},
\bauthor{\bsnm{Koumoutsakos}, \binits{P.}}:
\batitle{Data-driven forecasting of high-dimensional chaotic systems with long
  short-term memory networks}.
\bjtitle{Proceedings of the Royal Society A: Mathematical, Physical and
  Engineering Sciences}
\bvolume{474}(\bissue{2213}),
\bfpage{20170844}
(\byear{2018})
{\href{https://arxiv.org/abs/https://royalsocietypublishing.org/doi/pdf/10.1098/rspa.2017.0844}{{https://royalsocietypublishing.org/doi/pdf/10.1098/rspa.2017.0844}}}.
\doiurl{10.1098/rspa.2017.0844}
\end{barticle}
\endbibitem

%%% 15
\bibitem{Vlachas2022}
\begin{barticle}
\bauthor{\bsnm{Vlachas}, \binits{P.R.}},
\bauthor{\bsnm{Arampatzis}, \binits{G.}},
\bauthor{\bsnm{Uhler}, \binits{C.}},
\bauthor{\bsnm{Koumoutsakos}, \binits{P.}}:
\batitle{Multiscale simulations of complex systems by learning their effective
  dynamics}.
\bjtitle{Nature Machine Intelligence}
\bvolume{4}(\bissue{4}),
\bfpage{359}--\blpage{366}
(\byear{2022}).
\doiurl{10.1038/s42256-022-00464-w}
\end{barticle}
\endbibitem

%%% 16
\bibitem{Psarellis2022_2}
\begin{botherref}
\oauthor{\bsnm{Psarellis}, \binits{Y.M.}},
\oauthor{\bsnm{Lee}, \binits{S.}},
\oauthor{\bsnm{Bhattacharjee}, \binits{T.}},
\oauthor{\bsnm{Datta}, \binits{S.S.}},
\oauthor{\bsnm{Bello-Rivas}, \binits{J.M.}},
\oauthor{\bsnm{Kevrekidis}, \binits{I.G.}}:
Data-driven Discovery of Chemotactic Migration of Bacteria via Machine
  Learning.
arXiv
(2022).
\doiurl{10.48550/ARXIV.2208.11853}.
\url{https://arxiv.org/abs/2208.11853}
\end{botherref}
\endbibitem

%%% 17
\bibitem{Psarellis2022}
\begin{botherref}
\oauthor{\bsnm{Lee}, \binits{S.}},
\oauthor{\bsnm{Psarellis}, \binits{Y.M.}},
\oauthor{\bsnm{Siettos}, \binits{C.I.}},
\oauthor{\bsnm{Kevrekidis}, \binits{I.G.}}:
Learning black- and gray-box chemotactic PDEs/closures from agent based Monte
  Carlo simulation data.
arXiv
(2022).
\doiurl{10.48550/ARXIV.2205.13545}.
\url{https://arxiv.org/abs/2205.13545}
\end{botherref}
\endbibitem

%%% 18
\bibitem{Arbabi2021}
\begin{barticle}
\bauthor{\bsnm{Arbabi}, \binits{H.}},
\bauthor{\bsnm{Kevrekidis}, \binits{I.G.}}:
\batitle{Particles to partial differential equations parsimoniously}.
\bjtitle{Chaos: An Interdisciplinary Journal of Nonlinear Science}
\bvolume{31}(\bissue{3}),
\bfpage{033137}
(\byear{2021})
{\href{https://arxiv.org/abs/https://doi.org/10.1063/5.0037837}{{https://doi.org/10.1063/5.0037837}}}.
\doiurl{10.1063/5.0037837}
\end{barticle}
\endbibitem

%%% 19
\bibitem{Galaris2022}
\begin{barticle}
\bauthor{\bsnm{Galaris}, \binits{E.}},
\bauthor{\bsnm{Fabiani}, \binits{G.}},
\bauthor{\bsnm{Gallos}, \binits{I.}},
\bauthor{\bsnm{Kevrekidis}, \binits{I.}},
\bauthor{\bsnm{Siettos}, \binits{C.}}:
\batitle{Numerical bifurcation analysis of pdes from lattice boltzmann model
  simulations: a parsimonious machine learning approach}.
\bjtitle{Journal of Scientific Computing}
\bvolume{92}(\bissue{2}),
\bfpage{34}
(\byear{2022}).
\doiurl{10.1007/s10915-022-01883-y}
\end{barticle}
\endbibitem

%%% 20
\bibitem{Lee2020}
\begin{barticle}
\bauthor{\bsnm{Lee}, \binits{S.}},
\bauthor{\bsnm{Kooshkbaghi}, \binits{M.}},
\bauthor{\bsnm{Spiliotis}, \binits{K.}},
\bauthor{\bsnm{Siettos}, \binits{C.I.}},
\bauthor{\bsnm{Kevrekidis}, \binits{I.G.}}:
\batitle{Coarse-scale pdes from fine-scale observations via machine learning}.
\bjtitle{Chaos: An Interdisciplinary Journal of Nonlinear Science}
\bvolume{30}(\bissue{1}),
\bfpage{013141}
(\byear{2020})
{\href{https://arxiv.org/abs/https://doi.org/10.1063/1.5126869}{{https://doi.org/10.1063/1.5126869}}}.
\doiurl{10.1063/1.5126869}
\end{barticle}
\endbibitem

%%% 21
\bibitem{Duraisamy2015}
\begin{botherref}
\oauthor{\bsnm{Zhang}, \binits{Z.J.}},
\oauthor{\bsnm{Duraisamy}, \binits{K.}}:
Machine Learning Methods for Data-Driven Turbulence Modeling.
\doiurl{10.2514/6.2015-2460}.
\url{https://arc.aiaa.org/doi/abs/10.2514/6.2015-2460}
\end{botherref}
\endbibitem

%%% 22
\bibitem{Duraisamy2019}
\begin{barticle}
\bauthor{\bsnm{Duraisamy}, \binits{K.}},
\bauthor{\bsnm{Iaccarino}, \binits{G.}},
\bauthor{\bsnm{Xiao}, \binits{H.}}:
\batitle{Turbulence modeling in the age of data}.
\bjtitle{Annual Review of Fluid Mechanics}
\bvolume{51}(\bissue{1}),
\bfpage{357}--\blpage{377}
(\byear{2019})
{\href{https://arxiv.org/abs/https://doi.org/10.1146/annurev-fluid-010518-040547}{{https://doi.org/10.1146/annurev-fluid-010518-040547}}}.
\doiurl{10.1146/annurev-fluid-010518-040547}
\end{barticle}
\endbibitem

%%% 23
\bibitem{Lee2019}
\begin{barticle}
\bauthor{\bsnm{Lee}, \binits{S.}},
\bauthor{\bsnm{Dietrich}, \binits{F.}},
\bauthor{\bsnm{Karniadakis}, \binits{G.E.}},
\bauthor{\bsnm{Kevrekidis}, \binits{I.G.}}:
\batitle{Linking gaussian process regression with data-driven manifold
  embeddings for nonlinear data fusion}.
\bjtitle{Interface Focus}
\bvolume{9}(\bissue{3}),
\bfpage{20180083}
(\byear{2019})
{\href{https://arxiv.org/abs/https://royalsocietypublishing.org/doi/pdf/10.1098/rsfs.2018.0083}{{https://royalsocietypublishing.org/doi/pdf/10.1098/rsfs.2018.0083}}}.
\doiurl{10.1098/rsfs.2018.0083}
\end{barticle}
\endbibitem

%%% 24
\bibitem{Kemeth2022}
\begin{botherref}
\oauthor{\bsnm{Kemeth}, \binits{F.P.}},
\oauthor{\bsnm{Alonso}, \binits{S.}},
\oauthor{\bsnm{Echebarria}, \binits{B.}},
\oauthor{\bsnm{Moldenhawer}, \binits{T.}},
\oauthor{\bsnm{Beta}, \binits{C.}},
\oauthor{\bsnm{Kevrekidis}, \binits{I.G.}}:
Black and Gray Box Learning of Amplitude Equations: Application to Phase Field
  Systems.
arXiv
(2022).
\doiurl{10.48550/ARXIV.2207.03954}.
\url{https://arxiv.org/abs/2207.03954}
\end{botherref}
\endbibitem

%%% 25
\bibitem{Pan2018}
\begin{barticle}
\bauthor{\bsnm{Pan}, \binits{S.}},
\bauthor{\bsnm{Duraisamy}, \binits{K.}}:
\batitle{Data-driven discovery of closure models}.
\bjtitle{SIAM Journal on Applied Dynamical Systems}
\bvolume{17}(\bissue{4}),
\bfpage{2381}--\blpage{2413}
(\byear{2018})
{\href{https://arxiv.org/abs/https://doi.org/10.1137/18M1177263}{{https://doi.org/10.1137/18M1177263}}}.
\doiurl{10.1137/18M1177263}
\end{barticle}
\endbibitem

%%% 26
\bibitem{PARISH2016758}
\begin{barticle}
\bauthor{\bsnm{Parish}, \binits{E.J.}},
\bauthor{\bsnm{Duraisamy}, \binits{K.}}:
\batitle{A paradigm for data-driven predictive modeling using field inversion
  and machine learning}.
\bjtitle{Journal of Computational Physics}
\bvolume{305},
\bfpage{758}--\blpage{774}
(\byear{2016}).
\doiurl{10.1016/j.jcp.2015.11.012}
\end{barticle}
\endbibitem

%%% 27
\bibitem{Raissi2017}
\begin{botherref}
\oauthor{\bsnm{Raissi}, \binits{M.}},
\oauthor{\bsnm{Karniadakis}, \binits{G.}}:
Hidden physics models: Machine learning of nonlinear partial differential
  equations.
Journal of Computational Physics
\textbf{357}
(2017).
\doiurl{10.1016/j.jcp.2017.11.039}
\end{botherref}
\endbibitem

%%% 28
\bibitem{linot2023stabilized}
\begin{barticle}
\bauthor{\bsnm{Linot}, \binits{A.J.}},
\bauthor{\bsnm{Burby}, \binits{J.W.}},
\bauthor{\bsnm{Tang}, \binits{Q.}},
\bauthor{\bsnm{Balaprakash}, \binits{P.}},
\bauthor{\bsnm{Graham}, \binits{M.D.}},
\bauthor{\bsnm{Maulik}, \binits{R.}}:
\batitle{Stabilized neural ordinary differential equations for long-time
  forecasting of dynamical systems}.
\bjtitle{Journal of Computational Physics}
\bvolume{474},
\bfpage{111838}
(\byear{2023})
\end{barticle}
\endbibitem

%%% 29
\bibitem{floryan2022data}
\begin{barticle}
\bauthor{\bsnm{Floryan}, \binits{D.}},
\bauthor{\bsnm{Graham}, \binits{M.D.}}:
\batitle{Data-driven discovery of intrinsic dynamics}.
\bjtitle{Nature Machine Intelligence}
\bvolume{4}(\bissue{12}),
\bfpage{1113}--\blpage{1120}
(\byear{2022})
\end{barticle}
\endbibitem

%%% 30
\bibitem{Kevrekidis1990}
\begin{barticle}
\bauthor{\bsnm{Kevrekidis}, \binits{I.G.}},
\bauthor{\bsnm{Nicolaenko}, \binits{B.}},
\bauthor{\bsnm{Scovel}, \binits{J.C.}}:
\batitle{Back in the saddle again: A computer assisted study of the
  kuramoto-sivashinsky equation}.
\bjtitle{SIAM Journal on Applied Mathematics}
\bvolume{50}(\bissue{3}),
\bfpage{760}--\blpage{790}
(\byear{1990})
\end{barticle}
\endbibitem

%%% 31
\bibitem{he2015deep}
\begin{botherref}
\oauthor{\bsnm{He}, \binits{K.}},
\oauthor{\bsnm{Zhang}, \binits{X.}},
\oauthor{\bsnm{Ren}, \binits{S.}},
\oauthor{\bsnm{Sun}, \binits{J.}}:
Deep residual learning.
Image Recognition
\textbf{7}
(2015)
\end{botherref}
\endbibitem

%%% 32
\bibitem{Takens1981}
\begin{bbook}
\bauthor{\bsnm{Takens}, \binits{F.}}:
In: \beditor{\bsnm{Rand}, \binits{D.}},
\beditor{\bsnm{Young}, \binits{L.-S.}} (eds.)
\bbtitle{Detecting strange attractors in turbulence},
pp. \bfpage{366}--\blpage{381}.
\bpublisher{Springer},
\blocation{Berlin, Heidelberg}
(\byear{1981}).
\doiurl{10.1007/BFb0091924}.
\burl{https://doi.org/10.1007/BFb0091924}
\end{bbook}
\endbibitem

%%% 33
\bibitem{Whitney1936}
\begin{barticle}
\bauthor{\bsnm{Whitney}, \binits{H.}}:
\batitle{Differentiable manifolds}.
\bjtitle{Annals of Mathematics}
\bvolume{37}(\bissue{3}),
\bfpage{645}--\blpage{680}
(\byear{1936})
\end{barticle}
\endbibitem

%%% 34
\bibitem{everson1995karhunen}
\begin{barticle}
\bauthor{\bsnm{Everson}, \binits{R.}},
\bauthor{\bsnm{Sirovich}, \binits{L.}}:
\batitle{Karhunen--loeve procedure for gappy data}.
\bjtitle{JOSA A}
\bvolume{12}(\bissue{8}),
\bfpage{1657}--\blpage{1664}
(\byear{1995})
\end{barticle}
\endbibitem

%%% 35
\bibitem{r19}
\begin{barticle}
\bauthor{\bsnm{Coifman}, \binits{R.R.}},
\bauthor{\bsnm{Lafon}, \binits{S.}}:
\batitle{Diffusion maps}.
\bjtitle{Applied and computational harmonic analysis}
\bvolume{21}(\bissue{1}),
\bfpage{5}--\blpage{30}
(\byear{2006})
\end{barticle}
\endbibitem

%%% 36
\bibitem{r20}
\begin{barticle}
\bauthor{\bsnm{Nadler}, \binits{B.}},
\bauthor{\bsnm{Lafon}, \binits{S.}},
\bauthor{\bsnm{Coifman}, \binits{R.R.}},
\bauthor{\bsnm{Kevrekidis}, \binits{I.G.}}:
\batitle{Diffusion maps, spectral clustering and reaction coordinates of
  dynamical systems}.
\bjtitle{Applied and Computational Harmonic Analysis}
\bvolume{21}(\bissue{1}),
\bfpage{113}--\blpage{127}
(\byear{2006})
\end{barticle}
\endbibitem

%%% 37
\bibitem{r21}
\begin{barticle}
\bauthor{\bsnm{Coifman}, \binits{R.R.}},
\bauthor{\bsnm{Kevrekidis}, \binits{I.G.}},
\bauthor{\bsnm{Lafon}, \binits{S.}},
\bauthor{\bsnm{Maggioni}, \binits{M.}},
\bauthor{\bsnm{Nadler}, \binits{B.}}:
\batitle{Diffusion maps, reduction coordinates, and low dimensional
  representation of stochastic systems}.
\bjtitle{Multiscale Modeling \& Simulation}
\bvolume{7}(\bissue{2}),
\bfpage{842}--\blpage{864}
(\byear{2008})
\end{barticle}
\endbibitem

%%% 38
\bibitem{oron2002nonlinear}
\begin{barticle}
\bauthor{\bsnm{Oron}, \binits{A.}},
\bauthor{\bsnm{Gottlieb}, \binits{O.}}:
\batitle{Nonlinear dynamics of temporally excited falling liquid films}.
\bjtitle{Physics of Fluids}
\bvolume{14}(\bissue{8}),
\bfpage{2622}--\blpage{2636}
(\byear{2002})
\end{barticle}
\endbibitem

%%% 39
\bibitem{shklyaev2017longwave}
\begin{bbook}
\bauthor{\bsnm{Shklyaev}, \binits{S.}},
\bauthor{\bsnm{Nepomnyashchy}, \binits{A.}}:
\bbtitle{Longwave instabilities and patterns in fluids}.
\bpublisher{New York, NY, USA: Birkhäuser}, \blocation{???}
(\byear{2017})
\end{bbook}
\endbibitem

%%% 40
\bibitem{chang1986nonlinear}
\begin{barticle}
\bauthor{\bsnm{Chang}, \binits{H.-C.}}:
\batitle{Nonlinear waves on liquid film surfaces—i. flooding in a vertical
  tube}.
\bjtitle{Chemical engineering science}
\bvolume{41}(\bissue{10}),
\bfpage{2463}--\blpage{2476}
(\byear{1986})
\end{barticle}
\endbibitem

%%% 41
\bibitem{shlang1982irregular}
\begin{barticle}
\bauthor{\bsnm{Shlang}, \binits{T.}},
\bauthor{\bsnm{Sivashinsky}, \binits{G.}}:
\batitle{Irregular flow of a liquid film down a vertical column}.
\bjtitle{Journal de Physique}
\bvolume{43}(\bissue{3}),
\bfpage{459}--\blpage{466}
(\byear{1982})
\end{barticle}
\endbibitem

%%% 42
\bibitem{brown1992}
\begin{botherref}
\oauthor{\bsnm{Brown}, \binits{H.S.}}:
A computer-assisted, nonlinear dynamic study of instabilities and pattern
  formation for interfacial waves.
PhD thesis,
Princeton University
(1992)
\end{botherref}
\endbibitem

%%% 43
\bibitem{r25}
\begin{barticle}
\bauthor{\bsnm{Dsilva}, \binits{C.J.}},
\bauthor{\bsnm{Talmon}, \binits{R.}},
\bauthor{\bsnm{Coifman}, \binits{R.R.}},
\bauthor{\bsnm{Kevrekidis}, \binits{I.G.}}:
\batitle{Parsimonious representation of nonlinear dynamical systems through
  manifold learning: A chemotaxis case study}.
\bjtitle{Applied and Computational Harmonic Analysis}
\bvolume{44}(\bissue{3}),
\bfpage{759}--\blpage{773}
(\byear{2018})
\end{barticle}
\endbibitem

\end{thebibliography}

\end{document}